\documentclass[10pt,a4paper]{article}

\usepackage{latexsym}
\usepackage{graphics}
\usepackage{times}
\usepackage{cite}
\usepackage{upgreek}

\usepackage[left=2cm,right=2cm,top=2cm,bottom=2cm]{geometry}
\usepackage[font=normalsize]{caption}
\usepackage{authblk}

\begin{document}

\title{ \LARGE Ultracold neutron detection with $^6$Li-doped glass scintillators \\ \large NANOSC: a fast ultracold neutron detector for the nEDM experiment at the Paul Scherrer Institute}

\author[1]{G.~Ban}
\author[2]{G.~Bison}
\author[3]{K.~Bodek}
\author[2]{Z.~Chowdhuri}
\author[4]{P.~Geltenbort}
\author[5]{W.C.~ Griffith}
\author[1,2]{V.~H\'elaine\thanks{Now at LPSC, Grenoble, France.}}
\author[2]{R.~Henneck\thanks{Now retired.}}
\author[6]{M.~Kasprzak}
\author[7]{Y.~Kermaidic}
\author[2,8]{K.~Kirch}
\author[2,8]{S.~Komposch}
\author[6]{P.A.~Koss}
\author[9]{A.~Kozela}
\author[8]{J.~Krempel}
\author[2]{B.~Lauss}
\author[1]{T.~Lefort\thanks{Corresponding author: lefort@lpccaen.in2p3.fr}}
\author[1]{Y.~Lemi\`ere}
\author[2]{A.~Mtchedlishvili}
\author[5]{M.~Musgrave}
\author[1]{O.~Naviliat-Cuncic\thanks{Now at Michigan State University, East-Lansing, MI, USA.}}
\author[8]{F.~M.~Piegsa}
\author[10]{E.~Pierre}
\author[7]{G.~Pignol}
\author[1]{G.~Qu\'em\'ener}
\author[8]{M.~Rawlik}
\author[2,8]{D.~Ries}
\author[7]{D.~Rebreyend}
\author[11]{S.~Roccia}
\author[1]{G.~Rogel\thanks{Now at CERAP, Cherbourg, France.}}
\author[2]{P.~Schmidt-Wellenburg}
\author[6]{N.~Severijns}
\author[6]{E.~Wursten}
\author[3]{J.~Zejma}
\author[2]{G.~Zsigmond}

\affil[1]{Normandie Univ, ENSICAEN, UNICAEN, CNRS/IN2P3, LPC Caen, 14000 Caen, France}
\affil[2]{Paul Scherrer Institute, CH-5232 Villigen-PSI, Switzerland}
\affil[3]{Marian Smoluchowski Institute of Physics, Jagiellonian University, 30-059 Cracow, Poland}
\affil[4]{Institut Laue-Langevin, Grenoble, France}
\affil[5]{Department of Physics and Astronomy, University of Sussex, Falmer, Brighton BN1 9QH, United Kingdom}
\affil[6]{Instituut voor Kernen Stralingsfysica, Katholieke~Universiteit~Leuven, B-3001 Leuven, Belgium}
\affil[7]{LPSC, Universit\'e Grenoble Alpes, CNRS/IN2P3, Grenoble, France}
\affil[8]{ETH Z\"urich, Institute for Particle Physics, CH-8093 Z\"urich, Switzerland}
\affil[9]{Henryk Niedwodnicza\'nski Institute for Nuclear Physics, 31-342 Cracow, Poland}
\affil[10]{RCNP, Osaka University, 10-1 Mihogaoka, Ibaraki, Osaka, 567-0047, Japan}
\affil[11]{CSNSM, Universit\'e Paris Sud, CNRS/IN2P3, Orsay campus, France}

\date{\normalsize Date: June 23, 2016}

\setcounter{Maxaffil}{0}
\renewcommand\Affilfont{\itshape\small}

\maketitle

\abstract{This paper summarizes the results from measurements aiming to characterize ultracold neutron detection with $^6$Li-doped glass scintillators. 
Single GS10 or GS20 scintillators, with a thickness of 100--200 $\upmu$m, fulfill the ultracold neutron detection requirements with an acceptable 
neutron-gamma discrimination. This discrimination is clearly improved with a stack of two scintillators: a $^6$Li-depleted glass bonded to a 
$^6$Li-enriched glass. The optical contact bonding is used between the scintillators in order to obtain a perfect optical contact. The scintillator's 
detection efficiency is similar to that of a $^3$He Strelkov gas detector. Coupled to a digital data acquisition system, counting rates up to a few
10$^5$ counts/s can be handled. A detector based on such a scintillator stack arrangement was built and has been used in the neutron electric dipole 
moment experiment at the
Paul Scherrer Institute since 2010. Its response for the regular runs of 
the neutron electric dipole moment experiment is presented. 
} 

\vspace{0.4cm} 


\vspace{0.4cm} 


\section{Introduction}
\label{intro}

\begin{sloppypar}
The advent of high intensity ultracold neutron (UCN) sources \cite{Lau12, Zim11, Tri00, Mas02, Sau04, Ser09} requires the development of new UCN detectors
able to handle high counting rates. During the last 40 years, the so-called ``Strelkov'' $^3$He gas detector was the most commonly 
used detector in the UCN field. However, this counter is rather slow (2 $\upmu$s pulse length) and the $^3$He shortage
may become a serious and expensive issue in the near future. Counters based on $^6$Li and $^{10}$B nuclei have been predominantly investigated 
as potential alternatives to $^3$He. The new detectors are based on gaseous detection \cite{Kle11, Sal12, Mor09}, 
solid state silicon detection \cite{Bak03, Las05, Lau11} or $^6$Li-doped glass scintillators \cite{Ban05, Ban09, Bla15, Got13, Osh11}.
The following features are usually sought after: a detection efficiency as high as possible,  the ability to handle counting rates up to 10$^6$ counts/s, 
high discrimination power between UCN and background (mostly gamma rays and thermal neutrons), low production costs, stable
operation and easy maintenance. Developments looking for position-sensitivity are also being pursued \cite{Lau11, Jak09, Kaw10, Bae11}. 
Detection efficiency is one of the most important features. The detector entrance 
window, if any, must have a low Fermi potential (neutron optical potential) and a high UCN transmission \cite{Mor09, Atc09}. The use of 
a conversion layer is not recommended \cite{Dia01} since it reduces the UCN detection efficiency by about 10--30$\%$ \cite{Kle11, Las05, 
Rog09, Jak09} due to the absorption of the charged particles produced by the neutron capture in the layer.
\end{sloppypar}

In this context, we have investigated UCN detection using $^6$Li-doped glass scintillators \cite{Ban05, Ban09} with the aim of building a  
detector for the neutron electric dipole moment (nEDM) experiment at the Paul Scherrer Institute (PSI) in Switzerland. These scintillators are fast 
with a scintillation decay time of $\tau$ = 70 ns and Fermi potentials ranging between 80 and 100 neV. The neutron converter ($^6$Li) is in 
the detection volume so no extra conversion 
layer is required. An acceptable neutron-gamma discrimination can be achieved \cite{Ban05, Ban09}. The study of these scintillators is also being pursued by 
other groups \cite{Got13, Osh11, Bla15}. 

This article summarizes the results from measurements performed first at Institute Laue Langevin (ILL) PF2 beam lines and later at PSI. It is organized as follows:
the main features of the glass scintillators are summarized in Section \ref{properties}; UCN detection with single scintillators is described for four different 
scintillators in Section \ref{single}; the scintillator's counting rate capability is investigated in Section \ref{rate}; properties of scintillator stacks are 
studied in Section \ref{stack}; finally, the main features of a new detector (NANOSC), currently used in the nEDM experiment at PSI, is presented in 
Section \ref{NANOSC}.    

\section{Main properties of $^6$Li-doped scintillators}
\label{properties}

\subsection{Neutron detection}
\label{neutron_det}

The neutron detection is based on the $^6$Li neutron capture reaction:

\begin{center}
n + $^6$Li $\rightarrow$ $^3$H (2.74 MeV) + $^4$He (2.05 MeV).
\end{center}

The energy released by the neutron capture amounts to 4.79 MeV. This enables an efficient discrimination between
the background (gamma and electronic noise) and the neutron contributions. The capture cross section is 940 barns for thermal neutrons.
Following the $1/v$ law, the capture cross section in the UCN velocity range (between 3 and 10 m/s) is of the order 
of a few 10$^5$ barns. This enables the use of $^6$Li-doped glass scintillators with thicknesses around 
100--200 $\upmu$m while maintening a high detection efficiency. As a result, the amount of energy released by gamma interactions 
is minimized and the neutron gamma discrimination is improved \cite{Ban09}. 

In the tests reported here, four types of glass scintillator designated by GS{\it x} were used: GS3, GS30, GS10 and GS20. The GS30, GS10 and  
GS20 have the same chemical composition but differ in their lithium content whereas the chemical composition of  
GS3 is different \cite{AST}. The GS3 and the GS30 are depleted in $^6$Li, the GS20 is enriched 
in $^6$Li and the GS10 has the natural $^6$Li fraction as shown in the first line of table \ref{tab:tab_properties}.

\begin{table}[!htb]
\begin{center}
\captionsetup{width=0.8\textwidth}
\caption{\small Main properties of the $^6$Li-doped glass scintillators. The Fermi potential is calculated using the 
chemical composition of the glass scintillators \cite{AST} and tabulated data of the neutron scattering lengths \cite{PotFermi}. 
The critical velocity is derived from the calculated Fermi potential. The UCN mean free path is computed assuming a 
neutron kinetic energy of 130 neV (minus the Fermi potential of the corresponding scintillator).} 

\vspace{0.35cm}
\begin{tabular}{ccccc}
\hline
\hline
GS type & GS3 & GS30 & GS10 & GS20   \\
\hline
\footnotesize $^6$Li fraction ($\%$) & 0.01 & 0.01 & 7.5 & 95 \\ 
\footnotesize$^6$Li-density (cm$^{-3}$) & 6.4 10$^{17}$ & 1.8 10$^{18}$ & 1.4 10$^{21}$ & 1.8 10$^{22}$ \\
\footnotesize Fermi pot. (neV) & 102.0 &  83.1 &  84.6 & 103.4 \\ 
\footnotesize Critical vel. (m/s)  &  4.4  & 4.0 &  4.0  & 4.4  \\
\footnotesize Mean path ($\upmu$m) & 1.8 10$^4$&  0.8 10$^4$& 10.3  & 0.6 \\
\hline
\hline
\end{tabular}
\label{tab:tab_properties}
\end{center}
\end{table}

On the one hand, the GS3 and the GS30 scintillators, with a UCN mean free path of a few mm, are nearly transparent to UCN. On the other hand, 
the UCN mean free path for the GS10 and the GS20 are of the order of 1--10 $\upmu$m favoring their use for UCN detection. 
The Fermi potential of all scintillators allows UCN detection with energies above 84 neV or 103 neV depending on the 
scintillators' composition. Cold neutron reflectivity measurements performed at PSI confirmed the 
calculated value of the Fermi potential for GS10 scintillators.    

\subsection{Scintillation properties and features}
\label{scint_properties}

The scintillation materials obtained from Applied Scintillators Technology (AST) Ltd. and contain 
Ce$_2$O$_3$, SiO$_2$, MgO, Al$_2$O$_3$ and Li$_2$O oxides. Their mean densities are in the range of 
2.42--2.66 g cm$^{-3}$ and their refractive indexes range from 1.55 up to 1.58 \cite{AST}. They are robust and 
resistant to a wide range of organic and inorganic chemicals.

In the $^6$Li-doped glasses scintillators, the scintillation process is activated by Ce$^{3+}$ sites. The light emission peaks
in the blue region, at 395 nm. This allows the use of a wide variety of photomultiplier tubes without wavelength shifters. 
The scintillators are transparent to their own scintillation light and the decay time is in the range 60--75 ns 
\cite{Dia01}. As a result, counting rates up to a few 10$^5$ counts/s can be reached. On the other hand, the decay 
time is identical for neutrons and gammas, preventing any neutron-gamma discrimination based on a pulse shape analysis.
Therefore the discrimination is based on a pulse height (or charge) analysis. The number of photons per neutron capture is quoted at 
about 6000 photons 
while a 1 MeV gamma produces around 4000 photons \cite{Eijk01}. More details about the light output can be found in Ref. \cite{AST}.

\section{UCN detection with single scintillators}
\label{single}

\subsection{Pulse height distribution}
\label{pulse_height}

\begin{figure}
\begin{center}
\resizebox{0.4\textwidth}{!}
{\includegraphics{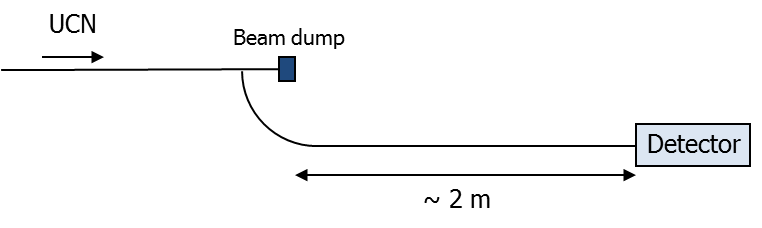}}
\captionsetup{width=0.8\textwidth}
\caption{\small Setup used for the scintillators tests at the ILL PF2 beam lines. The bent guide with a T shape is located just 
in front of the beam dump.}
\label{fig:beamline}
\end{center}
\end{figure}

\begin{figure}
\begin{center}
\resizebox{0.4\textwidth}{!}
{\includegraphics{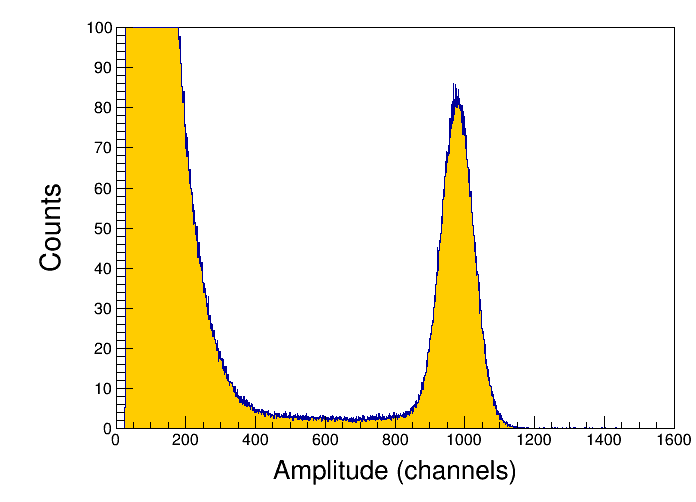}}
\captionsetup{width=0.8\textwidth}
\caption{\small Pulse height distribution measured with a 100 $\upmu$m thick $^6$Li-depleted scintillator (GS3). See text for details.}
\label{fig:singlescinti_GS3}
\end{center}
\end{figure}

Single scintillators were tested at the ILL PF2/EDM beam line. The direct fast UCN velocity 
component was suppressed using upstream a bent guide with a T shape (fig. \ref{fig:beamline}).
Three types of scintillators, the $^6$Li-depleted scintillators (GS3), the natural $^6$Li content scintillators 
(GS10) and the  $^6$Li-enriched scintillators (GS20), were studied. The scintillators had a diameter 
of 1 inch and a thickness of 100 $\upmu$m. The two scintillator faces were polished. In order to maximize the light 
collection, they were placed directly on a three inch PMT entrance window in front of the photocathode center (photonis XP53X2). Optical 
grease (Saint-Gobain BC-630) was used at the interface between the glass scintillators and the PMT
except for the GS3 scintillator (fig. \ref{fig:singlescinti_GS3}). Pulses were 
amplified by an Ortec 570 amplifier and their amplitudes were digitized with an Ortec ADC. The neutron counting 
rate was of the order of 1.5 $\times 10^3$ neutron/s for the GS10 and GS20 measurements. 

\begin{figure}
\begin{center}
\resizebox{0.4\textwidth}{!}
{\includegraphics{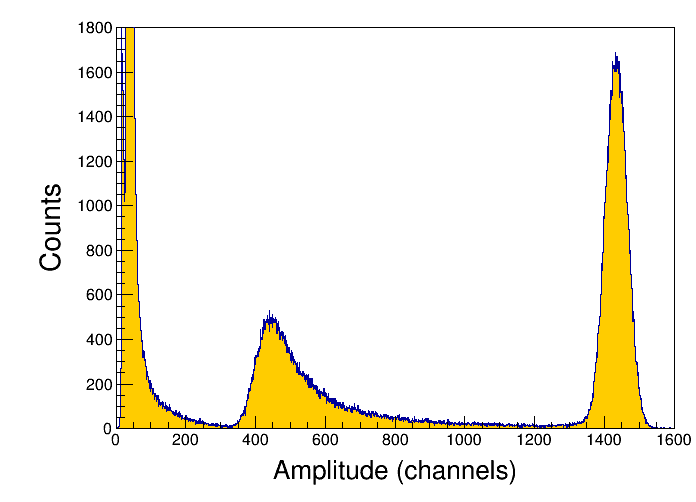}}
\captionsetup{width=0.8\textwidth}
\caption{\small Pulse height distribution measured with a 100 $\upmu$m thick $^6$Li-enriched scintillator (GS20). See text for details.}
\label{fig:singlescinti_GS20}
\end{center}
\end{figure}

\begin{figure}
\begin{center}
\resizebox{0.4\textwidth}{!}
{\includegraphics{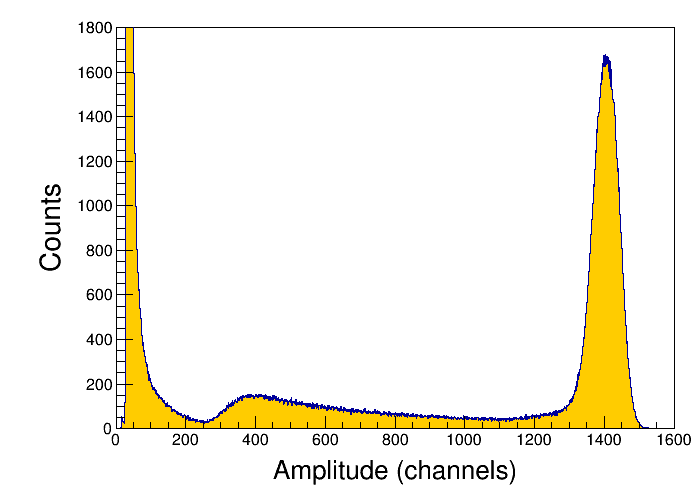}}
\captionsetup{width=0.8\textwidth}
\caption{\small Pulse height distribution measured with a 100 $\upmu$m thick natural $^6$Li content scintillator (GS10). See text for details.}
\label{fig:singlescinti_GS10}
\end{center}
\end{figure}

The pulse height distributions are shown in Figs. \ref{fig:singlescinti_GS20}, \ref{fig:singlescinti_GS10} and \ref{fig:singlescinti_GS3}.
Three contributions are visible: the electronic noise and gamma interactions at low amplitudes, the edge 
events in the intermediate range and the full-energy peak at large amplitudes. The full-energy peak corresponds to 
neutron captures taking place in the scintillator bulk: the alpha particles and tritons are stopped within the glass 
and their energy is fully deposited in the scintillator. The edge events are induced by neutron captures occurring close 
to the scintillator's entrance window. The alpha particle or the triton escapes from the glass and 
only a fraction of the neutron capture energy is deposited in the scintillator. The energy 
deposition of the gamma interaction is small because of the low thicknesses (100 $\upmu$m) of the scintillators. A detailed study 
of the gamma interactions within such scintillators was already reported in Ref. \cite{Ban09}. 

The amount of edge events depends on the UCN mean free path (Table \ref{tab:tab_properties}) and 
the triton and alpha particle ranges, being equal to respectively 36 $\upmu$m and 7 $\upmu$m for the studied scintillators \cite{SRIM}. 
Since the UCN mean free path mostly depends on the scintillators' $^6$Li content (and slightly on the UCN energy spectrum), the edge 
fraction varies from one scintillator type to another. For the tests 
reported here, 43$\%$ of the UCN interactions in the $^6$Li-enriched scintillator (GS20) lead to edge events, while this fraction 
decreases to 31$\%$ in the natural $^6$Li content scintillators (GS10). The main uncertainty comes from the measurement 
reproducibility which is at the 2$\%$ level. The shape of the edge events distribution is slightly different for the GS10 and the GS20. 
For the GS20, the UCN mean free path is much shorter than the particle ranges. Therefore, the escaping triton or alpha particle deposits
little energy in the glass and the edge events distribution is mainly located at low charge close to the gamma distribution. It 
mainly corresponds to the energy of the trapped ionizing particle (mostly the alpha particle since the triton range is 5 times larger). On the 
other hand, the escaping particle has a longer path in the GS10 scintillator since the neutron capture takes place at a deeper location in the 
scintillator. This causes the edge event distribution to be more spread out (fig. \ref{fig:singlescinti_GS10}).   
 
For the $^6$Li-depleted scintillator (GS3), there are nearly no edge events as observed in fig. \ref{fig:singlescinti_GS3}. For such a scintillator, 
the gamma interaction is the main contribution and the UCN counting rate corresponds to 4.5$\%$ of the total UCN 
counting rate measured with either the GS10 or the GS20. A similar distribution was measured for the GS30 $^6$Li-depleted 
scintillator. Although a larger UCN counting rate was expected because of the $^6$Li content being three times larger (
Table \ref{tab:tab_properties}) and a lower critical velocity, the UCN counting rate was halved. There is no clear reason for such a 
behaviour except that the $^6$Li content might not be as stated by the manufacturer.

\subsection{Detection efficiency}
\label{single_deteff}

A digital data acquisition system was developed at LPC Caen and was used for all measurements reported hereafter. 
The system, FASTER for Fast Acquisition SysTem for nuclEar Research, is based on FPGA (Field Programmable Gate Arrays) technology \cite{Fas15}. 
The FASTER hardware is made of mother boards (SYROCO) 
which host two daughter boards (CARAS). Each daughter board has two analog-to-digital converters able to sample pulses 
at a rate of 500 MHz with 12-bit resolution. The digitized pulses are then transferred to the first FPGA where a firmware programmed in 
VHDL (VHSIC (Very High Speed Integrated Circuit) Hardware Description Language)
carries out signal treatment (filtering and base line restoration), triggering (7.8 ps resolution for the digital constant fraction discriminator) and charge measurement. 
In addition, every event is time-stamped with a single clock (2 ns resolution) common to all channels. A second FPGA gathers the computed 
charges from both daughter boards and performs the transfer to a computer through a 1Gbit/s ethernet link. Finally, the system is 
driven with a custom-made acquisition software which uses ROOT for histogram building and visualization \cite{Bru97}. 

The pulse charge is measured instead of the amplitude in order to efficiently reject \v{C}erenkov events. Indeed, for scintillators coupled 
to light guides, gamma interactions produce electrons which generate \v{C}erenkov light mainly in the light guide. Such \v{C}erenkov events exhibit 
pulse heights of the same order of magnitude as the neutron pulse height. On the other hand, the width of the \v{C}erenkov pulses is a few ns 
at half-height compared with a few tens of ns for the neutron pulses. As a result, the charge measurement enables an improved discrimination between the 
\v{C}erenkov and the neutron contributions and guarantees a reliable measurement of the neutron counting rate. The pulse charge was computed 
with an integration gate of 400 ns.   

\subsubsection{Comparison between GS10 and GS20}

The relative detection efficiency between the natural $^6$Li content scintillator (GS10) and the $^6$Li-enriched 
scintillator (GS20) was studied. The set-up is described in Section \ref{pulse_height}. The tests were performed at the PF2/TES beam line. Both 
scintillators had a diameter of 25.4 mm and a thickness of 100 $\upmu$m. The gamma and the \v{C}erenkov interactions were discriminated 
from neutron interactions using a charge threshold defined by the location of the minimum between the background contribution and the 
edge events contribution around channel 250 in fig. \ref{fig:singlescinti_GS10}. 

Although the critical velocity for GS20 is higher than for GS10 (Table \ref{tab:tab_properties}), 
the same UCN counting rates were measured for both scintillators within the uncertainty of about 2$\%$. It turned 
out that the GS10 scintillator is too thin to fully stop fast UCN. This was studied using a 
GS10-GS20 stack with a GS20 scintillator placed behind the GS10 scintillator. The 
UCN counting rate increased by about 5--8$\%$ with respect to the single GS10 scintillator rate. On the other hand, no 
counting rate increase was measured with a GS20-GS20 stack. These measurements show that a thickness of 100 
$\upmu$m is enough for the GS20 to fully stop the UCN flux while a thickness in the 150--200 $\upmu$m range is needed for
the GS10.

\subsubsection{Comparison between GS10 scintillators and a $^3$He gas detector}
\label{comp_GS10_He3}
The detection efficiency of the scintillators relative to a $^3$He gas detector was investigated for
natural $^6$Li content scintillators (GS10). The counting rates of three GS10 scintillators with different 
thicknesses, 100 $\upmu$m, 250 $\upmu$m and 500 $\upmu$m, were compared to the counting rate of a $^3$He 
gas detector (Strelkov type). The measurements were performed at the PF2/EDM beam line at ILL. 
A straight beam line was assembled between the PF2 turbine and the detectors. With such an arrangement, the UCN velocity distribution 
exhibits a maximum around 8 m/s and extends up to 20 m/s \cite{Rog09}. The $^3$He gas detector was a proportional 
counter containing 1100 mbar of argon, 15 mbar of $^3$He and 15 mbar of carbon dioxide.
The 250 $\upmu$m thick GS10 was square in shape (30 $\times$ 30 mm$²$) and was glued to a 6 cm long 
light guide made of PolyMethylMethAcrylate (PMMA). The light readout was performed with a Photonis XP3112 PM tube. Optical grease was used between 
the light guide and the PM tube. The 100 $\upmu$m and 
500 $\upmu$m thick GS10 had a 76 mm diameter and were mounted directly on the PMT's entrance window in front of the photocathode center (Photonis XP53X2B). 
No optical grease was used. A 5 mm thick polyethylene collimator with a central circular aperture of 8 mm was mounted 
in front of the detector position, where all four detectors were interchanged. The pulses' charges were measured for the 
scintillator whereas the amplitudes were recorded for the $^3$He gas detector.

\begin{table}[!htb]
\begin{center}
\captionsetup{width=0.8\textwidth}
\caption{\small Detection efficiencies of GS10 scintillators relative to a Strelkov $^3$He gas detector 
measured at the PF2/EDM beam line at ILL. The fast UCN component was not removed for this set of 
measurements (see text for details).} 
\vspace{0.35cm}
\begin{tabular}{cccc}
\hline
\hline
GS10 thickness &  100 $\upmu$m & 250 $\upmu$m & 500 $\upmu$m  \\
\hline
Relative efficiency ($\%$) & 84.2  $\pm$ 0.2 & 94.8 $\pm$ 2.6  & 120.5 $\pm$ 3.5 \\ 
\hline
\hline
\end{tabular}
\label{tab:GS10_releff}
\end{center}
\end{table}

The GS10 detection efficiency increases with scintillator thickness (Table \ref{tab:GS10_releff}). The 100 $\upmu$m thick GS10 
scintillator is too thin to 
fully stop the fast UCN fraction. The velocity distribution ranges up to 20 m/s and the mean free path of a UCN with a velocity 
of 8 m/s is already 16.5 $\upmu$m in GS10. The scintillator with a thickness of 250 $\upmu$m has a similar detection efficiency 
compared to the $^3$He gas detector. On the other hand, the thickest GS10 scintillator is more efficient. As the gas detector 
is not able to fully stop the fastest UCN. The UCN mean free path with a velocity of 8 m/s is equal to 1.5 cm in the gas while the 
detection volume depth is of the order of 6 cm.   

The uncertainty reported in table \ref{tab:GS10_releff} accounts for the procedure used for the neutron-gamma 
discrimination. A charge threshold is used to distinguish between the gamma distribution, located at low charge, 
and the neutron distribution, lying at higher charge. This threshold cannot be easily defined when the charge 
distribution does not exhibit a minimum between both contributions (Ref. \cite{Ban09}). 
This is valid for the two thickest GS10 scintillators, for which the gamma distribution tail extends into the 
region of the edge events distribution. As a result, two scintillator counting rates were estimated: one with a 
conservative (high) threshold and one with a permissive (low) threshold. The relative efficiency reported in Table 
\ref{tab:GS10_releff} is computed with the mean rate and the error corresponds to the average between the maximum rate 
and the minimum rate. In addition, the measurements reproducibility was estimated by unmounting and 
remounting the detectors on the beam line. The corresponding uncertainty is at most 2.5 $\%$ and has to be 
taken into account as well in table \ref{tab:GS10_releff}.  

\section{Counting rate capability}
\label{rate}

The ability to handle high counting rates was investigated for rates up to 10$^6$ counts/s, first with a photomultiplier alone, 
then with a GS10 scintillator coupled to the PMT. At the same time, the performance of the FASTER acquisition system was tested.
 
The study of the PMT rate capability was performed with a blue light emitting diode flashing towards the PMT photocathode 
through a diffusing box. The diffusing box was required to suppress the dependence of the light intensity on the light direction 
which varies with the flashing frequency. An RC circuit, with a 15 ns time constant, was used in order to mimic the neutron pulse shape.
At the PMT location, the amount of light was of the same order of magnitude as the amount of light produced by  
a neutron capture. The diode was triggered by an RF generator with an adjustable frequency
ranging from 10 kHz up to 1 MHz. No change of the PMT gain was observed up to a frequency of 1 MHz. Higher 
rates were not tested.

\begin{sloppypar}
The scintillator tests were performed at PSI at the SINQ/BOA cold neutron beam line where fluxes 
up to 6 $\times$ 10$^7$ n/cm$^2$/s were available \cite{Mor14}. A 25$\times$25 mm$^2$ square piece of GS10 with a thickness of 
1 mm was used. The scintillator thickness was increased in order to adapt the detector efficiency to cold 
neutron energies. The GS10 was mounted on a prismatic PMMA light guide designed with a right-angle in order to put 
the photomultiplier out of the direct beam line-of-sight. The PMT was a square and compact R11187 Hamamatsu tube with a transit time of 
7 ns. The light guide and the scintillator were wrapped with a few layers of PolyTetraFluoroEthylene (PTFE) and aluminum foil. Optical grease was 
applied between the GS10, the light guide and the PMT. The cold neutron flux was varied using a set of $^6$Li 
polymer collimators located upstream and in front of the scintillator. 
\end{sloppypar}

\begin{figure}
\begin{center}
\resizebox{0.4\textwidth}{!}
{\includegraphics{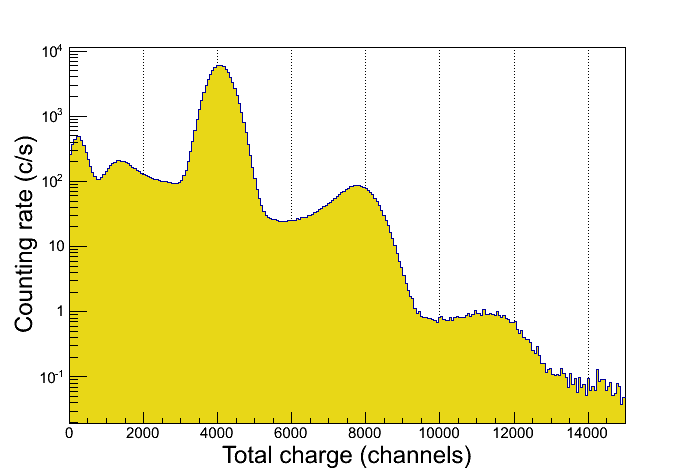}}
\captionsetup{width=0.8\textwidth}
\caption{\small Charge distribution measured with a 1 mm thick natural $^6$Li content scintillator (GS10) at the SINQ/BOA 
cold neutrons beam line at PSI. The counting rate was 9.2$\times$10$^4$ counts/s.}
\label{fig:BOASINQ}
\end{center}
\end{figure}

A charge distribution is shown in fig. \ref{fig:BOASINQ} corresponding to a counting rate of 9.2$\times$10$^4$ 
counts/s. The peak observed in the low charge region corresponds to electronic noise and gamma interactions which induced 
\v{C}erenkov events within the light guides (Section \ref{comp_GS10_He3}). The bump measured around channel 1400 
corresponds to gamma interactions in the scintillator and few edge events. Indeed, the 1 mm thickness of the GS10 increases the gamma 
interaction probability within the  scintillator as well as the amount of deposited energy \cite{Ban09}. Finally, the high counting 
rates induce pile-up events. For such pile-up events, the full-energy 
peak location is basically a multiple of the single neutron peak location as seen in fig. \ref{fig:BOASINQ}. For instance,
the simultaneous detection of two neutrons (around channel 8000) corresponds to twice the charge of the single neutron detection (around 
channel 4000). All these features explain why the charge spectrum is significantly different to the charge distribution 
measured with a 100 $\upmu$m thick GS10 scintillator in a moderate UCN flux without light guides (fig. \ref{fig:singlescinti_GS10}).

Counting rates from 10$^4$ up to 10$^6$ counts/s were measured. The FASTER acquisition system presented
a dead time of 80 ns per pulse resulting in a global dead time of 8$\%$ at 10$^6$ counts/s. At high rates, the amount 
of data that can be transferred between 
the front-end board and the computer is the limiting factor. The system was able to handle 
continuous rates up to 4$\times$10$^5$ counts/s without any losses. Above this rates, a fraction of the data was not
written to the disc. The charge associated with the neutron full-energy peak was unchanged up to 
a rate of 2$\times10^5$ counts/s. Above this value, a slight shift towards larger charges was observed (15 \% from
1.9$\times$10$^4$ to  5.6$\times$10$^5$ counts/s). This effect is attributed to a delayed emission of photons
 with a typical decay time of the order of $\upmu$s. At high counting rates, the pileup of such 
tail shifts measured charges at larger values. The origin of such a delayed emission is not fully understood. However, even 
at the largest rate, 10$^6$ count/s, neutron detection 
could still be achieved. When repeating measurements at low counting rates, no change in the charge distribution was observed as already 
noticed in Ref. \cite{Ban05} where no aging effect is observed with an absorbed number of neutrons of $10^{13}$ cm$^{-3}$.  

\subsection{Summary}

Single GS10 and GS20 scintillators are well suited for UCN detection \cite{Ban05} while GS3 and GS30 glasses are 
nearly transparent to UCN. The thicknesses of GS10 and GS20 have to be as low as possible in order to minimize 
the contributions from the gamma and thermal neutrons but large enough to fully stop UCN. The high $^6$Li enrichment of 
GS20 allows the use of scintillators thinner than the GS10 scintillators. Full UCN detection can be achieved with thicknesses 
lower than 100 $\upmu$m (although mechanical fragility then becomes an issue). On the other hand, the critical velocity and 
the amount of edge events are larger for the GS20 than for the GS10. The latter feature is a serious 
drawback when the discrimination between the edge events and the gamma contributions is blurred by the use of light guides 
which decrease the amount of collected light. For the GS10 scintillators, a thickness between 150 $\upmu$m and 200 $\upmu$m is large enough 
to fully stop UCN. Their critical velocity as well as their amount of edge events are smaller favoring their choice for UCN 
detection with a single scintillator.   
The GS10 detection efficiency is similar to the gas detector efficiency as long as its thickness is large enough
to fully stop the fastest UCN.   
Coupled to a R11187 Hamamatsu PMT, the scintillators are able to handle counting rates up to 10$^6$ counts/s
although a specific treatment of pile-up events is necessary.

\section{UCN detection with a stack of scintillators}
\label{stack}

\subsection{Principle and charge distribution}
\label{stack_principle}
\begin{figure}
\begin{center}
\resizebox{0.35\textwidth}{!}
{\includegraphics{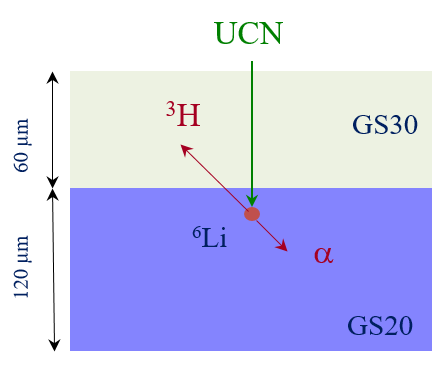}}
\captionsetup{width=0.8\textwidth}
\caption{\small Sketch of the scintillator stack with an edge event occurring close to the entrance window of the $^6$Li 
enriched scintillator.}
\label{fig:stack_principle}
\end{center}
\end{figure}

The neutron-gamma discrimination and the pile-up treatment can be further improved by suppressing the edge events. 
This was demonstrated with a stack of two scintillators consisting of a $^6$Li-depleted scintillator located in front of a $^6$Li 
enriched scintillator. A sketch of such an arrangement is shown in Fig \ref{fig:stack_principle}. According to 
Table \ref{tab:tab_properties}, 
the $^6$Li-depleted glass is nearly transparent to UCN whereas the enriched one stops all UCN. When an edge 
event takes place close to the entrance window of the $^6$Li-enriched scintillator, the escaping particle is travelling into 
the $^6$Li-depleted scintillator where it is stopped. As a result, the full-energy of the 
neutron capture is recovered and two well separated contributions can be distinguished: the gamma interactions, \v{C}erenkov events,
electronic noise at low charge, and the full-energy peak at large charge.  Ideally, the thickness of the first 
scintillator has to be slightly larger than the triton range (37 $\upmu$m), and the thickness of the 
second stage must be at least equal to five times the UCN mean free path in the $^6$Li-enriched scintillator. Both thicknesses 
have to be as small as possible in order to reduce the gamma interaction probability as well as the amount of deposited 
energy from the gamma interactions \cite{Ban09}.  

The test of a stack prototype was already reported in Ref. \cite{Ban09}. The edge events contribution 
was strongly reduced. However, due to a light-collection defect at the interface between the two glasses, 
a double peak structure was then observed instead of the expected single full-energy peak. 
Since optical grease or glue cannot be applied at the glasses interface without lowering the detection efficiency, 
the optical contact bonding was used. With this technique no substance is required: attractive forces 
(Van der Waals and/or covalent forces) strongly bond the touching faces and therefore ensure the 
optical contact. This holds when the contact surfaces are very clean and extremely flat, requiring a dedicated 
polishing procedure leading to a final mean roughness of the order of nm. 
The Soci\'et\'e Europ\'eenne de Syst\`emes Optiques (SESO) carried out this bonding method between 
GS30 and GS20 plates. The bonding was performed 
with 2 mm thick scintillators in order to be able to apply the necessary pressure to their external faces. Then 
the scintillator thicknesses were decreased by polishing, down to 60 $\upmu$m for the GS30 
and 120 $\upmu$m for the GS20. Further abrasion was not possible due to the fragility of glasses. 

\begin{figure}
\begin{center}
\resizebox{0.4\textwidth}{!}
{\includegraphics{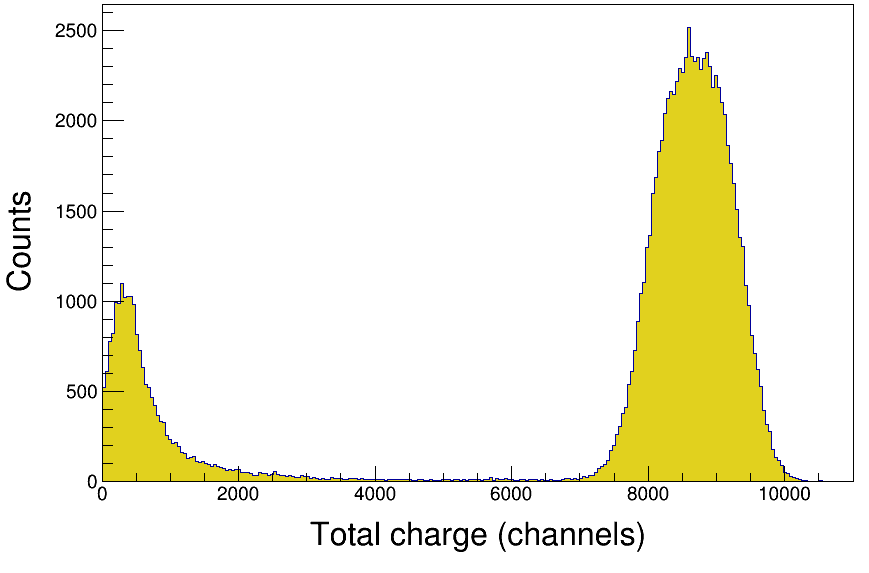}}
\captionsetup{width=0.8\textwidth}
\caption{\small Charge distribution measured with a GS30-GS20 stack using the optical contact bonding technique. The charge was integrated between 
0 and 300 ns.}
\label{fig:molstick}
\end{center}
\end{figure}

This scintillator stack was tested at the ILL PF2/TES beam line with the setup described in Section \ref{pulse_height}. 
The stack had a diameter of 25.4 mm. It was directly placed on the center of a 76.2 mm diameter   
PMT (Photonis XP3112). Optical grease (Saint-Gobain BC-630) was applied between the stack and the 
PMT. The charge distribution measured with the FASTER acquisition system is shown in fig. \ref{fig:molstick}. 
The charge distribution exhibits two contributions: a full-energy peak well separated from the low charge contribution.
Such a feature permits the improvement of the neutron-gamma discrimination with respect to a single scintillator.

\subsection{Search for the best stack combination}
\label{best_stack}

The search for the most efficient stack was performed with two $^6$Li-depleted scintillators for the entrance, 
a GS3 and a GS30, and two scintillators for the backing, a GS10 (natural $^6$Li content) 
and a GS20 (enriched in $^6$Li). For the first stage, different thicknesses were used. The UCN counting rate of the 
following combinations GS3-GS10, GS30-GS10, GS3-GS3-GS10, GS3-GS20 and GS30-GS20 were measured 
with the setup described in Section \ref{pulse_height} at the PF2/TES beam line at ILL. The optical contact bonding 
was not used, instead, scintillators were held together by a PTFE ring. All scintillators had a diameter of 25.4 mm, a 
thickness of 100 $\upmu$m and two polished faces. For each measurement, the detector block was dismounted and  
installed back again on the beam line with a new scintillator stack. The counting rate reproducibility of such an 
operation was measured to be around 2$\%$ while the statistical error was below 0.5$\%$. The neutron-gamma 
discrimination used a single threshold applied in the charge distribution. 

\begin{table}[!htb]
    \centering
    \captionsetup{width=0.8\textwidth}
    \caption[Detection efficiency for several scintillator stacks.]{Relative detection efficiencies measured for six scintillator stacks. 
The efficiencies presented in the table correspond to the counting rates measured by the scintillator stack divided by the 
counting rate measured by the single scintillator used as the backing of the stack. The asterisk indicates a GS30-GS20 
stack with a thickness of 60 $\upmu$m for the GS30 and 120 $\upmu$m for the GS20.}
       \vspace{0.35cm}
       \begin{tabular}{cccc}
        \hline
        \hline
        Stack &  \scriptsize GS30-GS10 & \scriptsize GS3-GS10 & \scriptsize GS3-GS3-GS10   \\
        Efficiency/\scriptsize GS10 & $80.6 \pm 2.3 \,\%$ & $72.2 \pm 2.1 \,\%$ & $54.2 \pm 1.6 \,\%$ \\
        \hline
        Stack & \scriptsize GS30-GS20$^{\ast}$ & \scriptsize GS30-GS20 & \scriptsize GS3-GS20 \\
       Efficiency/\scriptsize GS20 & $95.0 \pm 2.7 \,\%$ & $88.0 \pm 2.5 \,\%$ & $81.3 \pm 2.3 \,\%$ \\
        \hline
        \hline
        \end{tabular}
    \label{tab:releff}
\end{table}

In a first step, the UCN counting rates of the backing scintillator, {\it i.e.} the GS10 or GS20 scintillators, were 
measured. They were consistent within the error bars. Then, the stacks counting rates were measured and normalized 
to the counting rates measured with the backing scintillator. Results are summarized in 
Table \ref{tab:releff}. As expected, the stack detection efficiency decreases with the thickness of the first layer. This is clearly 
observed with the GS3-GS3-GS10 combination for which two 100 $\upmu$m thick GS3 scintillators were used in front of a GS10. 
This decrease is attributed to UCN scattering at the scintillator surface and n-gamma reactions since UCN capture in $^6$Li-depleted 
scintillators was measured at the level 
of a few $\%$ (Section \ref{pulse_height}). The GS30 scintillator had a 7 to 8$\%$ larger transmission than 
the GS3 scintillator whatever scintillator (GS10 or GS20) was used for the backing. The stacks 
made with GS20 for the backing were more efficient than the ones made with GS10. The counting rates of the GS3-GS20 and 
the GS30-GS20 were respectively 9$\%$ and 8$\%$ larger than the ones of GS3-GS10 and GS30-GS10.        

\subsection{Summary}
The optimal combination is the GS30-GS20 arrangement. The optical contact bonding was applied for this stack. The GS30 thickness 
was reduced to 60 $\upmu$m while the backing thickness 
was set to 120 $\upmu$m (Section \ref{stack_principle}). The stack efficiency relative to a single GS20
is reported in Table \ref{tab:releff} for the GS30-GS20 (marked with the asterisk). The reduction of the GS30 
thickness increases its transmission and results in the largest measured relative efficiency of 95$\%$ among the tested stacks.  

\section{Segmented detection system: NANOSC}
\label{NANOSC}

A new detector based on the GS30-GS20 scintillator stack was designed for the nEDM experiment at PSI \cite{Bak11}. The main 
goal was to build a detector with high rate capability and high detection efficiency. Its name, NANOSC, stands for NANO (nine) 
Scintillator Counters.

\subsection{Mechanical design}

\begin{figure}
\begin{center}
\resizebox{0.45\textwidth}{!}
{\includegraphics{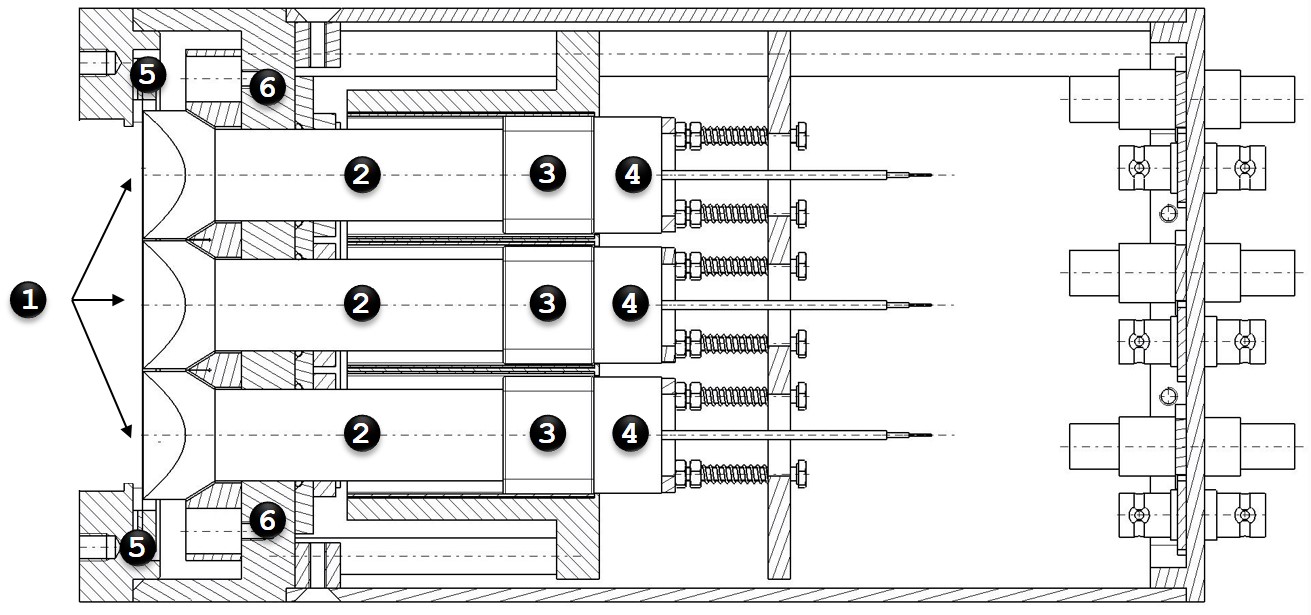}}
\captionsetup{width=0.8\textwidth}
\caption{\small Mechanical design of the NANOSC detector. \textbf{1:} scintillators stack, \textbf{2:} light guides, \textbf{3:}, PM tubes, \textbf{4:} voltage 
dividers, \textbf{5:} reflecting Al foil \textbf{6:} light guides holder and vacuum barrier}
\label{fig:NANOSC_design}
\end{center}
\end{figure}

\begin{figure}
\begin{center}
\resizebox{0.4\textwidth}{!}
{\includegraphics{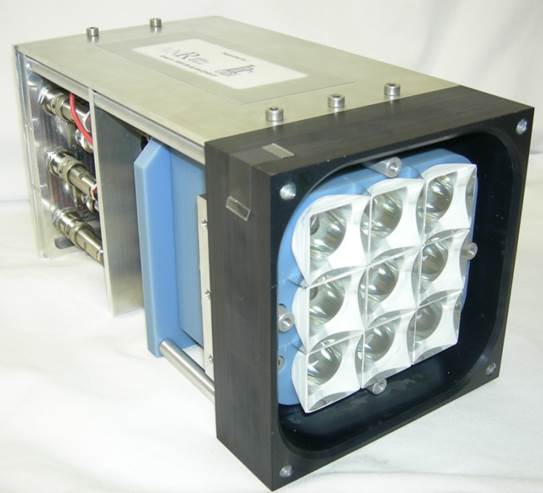}}
\captionsetup{width=0.8\textwidth}
\caption{\small Picture of the NANOSC front face showing the scintillator array. The blue piece seen on the side and in the front holds the nine light guides together. 
PTFE layers around the light guides and the scintillator sides are visible.}
\label{fig:NANOSC_picture}
\end{center}
\end{figure}

\begin{sloppypar}
The detector is made of nine independent channels. fig. \ref{fig:NANOSC_design} shows a side view of the detector's mechanical design. 
Each channel is composed of a square 28$\times$28 mm$^2$ GS30-GS20 scintillator stack coupled to an 
80 mm long PMMA light guide and a photomultiplier tube (R11187 Hamamatsu). The backing scintillator (GS20) is glued to the light 
guide while optical grease is used between the light guide and the PMT. The light guides are wrapped with two PTFE layers. 
The PMTs are in air while the scintillators are in vacuum. The vacuum tightness is accomplished by a plate with nine 
holes into which the nine light guides are inserted. O-rings provide the vacuum tightness. A single 10 $\upmu$m 
thick aluminum foil is placed in front of the nine scintillators in order to reflect the forward emitted scintillation light towards 
the PMTs. The amount of collected light is increased by 40--45$\%$ with this foil. 
The front detector box is blackened by anodization in order to prevent any extra and spurious reflected light 
(fig. \ref{fig:NANOSC_picture}). The total detector size is 100$\times$100$\times$300 mm$^3$ and its weight is 2 kg. 
\end{sloppypar}

In the nEDM experiment, the detector is installed below the spin analyzer which uses static and RF magnetic fields \cite{Hel15}. 
At the PMTs level, 
the stray field components are large enough to prevent standard PMTs to properly operate. For this reason, the compact Hamamatsu R11187 
PMT was selected. They are made of a square 18$\times$18 mm$^2$ Bialkali photocathode and a metal channel dynode system 
composed of 8 stages.  
This metal dynode structure presents a weak sensitivity to magnetic field and is mostly sensitive to the transverse 
magnetic field components 
with respect to the PMT's axis \cite{Ham}. In addition, every PMT is inserted in an individual mu-metal shield. 
Finally, the length of the light guides is 
80 mm in order to place the PMT in a
region where the transverse field components are below 2.5 mT, the limit for which PM gain variations vanish. Each PMT 
is wrapped in a Kapton$^{\textregistered}$ foil in order to suppress possible discharges between them. 

The UCN detection is then performed with a square array (86$\times$86) mm$^2$ of 3$\times$3 scintillator stacks as shown 
in Fig \ref{fig:NANOSC_picture}. 
The gaps between the nine scintillators amount for 5$\%$ of the total detection surface. Each 
channel is able to handle a counting rate of up to a few 10$^5$ counts/s leading to a total counting rate of a 
few 10$^6$ counts/s for the 
full detector. The signal treatment is performed with the FASTER acquisition system described in Section \ref{single_deteff}.

\subsection{Detection efficiency}
\label{NANOSC_efficiency}

The NANOSC detection efficiency was compared to the detection efficiency of $³$He gas detectors (Strelkov type). Measurements 
were performed at the PF2/TES beam line at ILL with three different experimental setups. For all measurements, the fast 
UCN component was suppressed either by using a chicane of 
stainless steel guides or by using the UCN velocity shaper described in Ref. \cite{Dau12}. 

The first set of measurements was performed with two different bent guides, one had a T shape, the other one an S shape. 
The "T-shape setup", shown in fig. \ref{fig:beamline}, suppressed the fast UCN velocity component more efficiently leading 
to a decrease of the UCN counting rate by a factor of 6.5. With this setup, a NANOSC detection efficiency of 87$\%$ 
relative to the gas detector was measured. On the other hand, the relative efficiency between the NANOSC detector and the gas detector 
went to 110$\%$ when the S-shape setup was used. The difference between the Fermi 
potential of both detectors entrance windows (103 neV for the GS scintillator instead of 54 neV for the gas detector) may explain 
the lower efficiency observed with the NANOSC detector when the T-shape setup was used.

The same trend was observed with measurements performed with the velocity shaper. The NANOSC detector efficiency was
95$\%$ relative to the gas detector when the detectors were placed exactly at the same location downstream of the 
shaper. On the other hand, the NANOSC detector became more efficient than the gas detector when it was placed at a lower height 
than the gas detector. Placing the NANOSC detector 60 cm lower than the beam height and 35 cm lower than the gas detector 
showed a relative efficiency of 130$\%$. The latter measurement demonstrated that when the difference of Fermi potential 
between the two detectors was counter-balanced by an equivalent change in height, the NANOSC detector was 
more efficient.

In the nEDM experiment, the detector is installed 212 cm below the UCN storage chamber. The minimum 
UCN energy (217 neV) is therefore well above the scintillator's Fermi potential ensuring a maximum detection efficiency.  

\subsection{Detector performance at the nEDM experiment}
\label{NANOSC_perf}

The NANOSC performance was characterized during the 2014 nEDM runs at PSI. These investigations are important since the number of detected 
UCN, more precisely the number of spin-up and spin-down UCN, are used in the first step of the nEDM analysis in order to extract the neutrons' frequency 
\cite{Bak13}. Since 2014, a simultaneous spin analyzer is installed below the nEDM spectrometer \cite{Hel15}. The device splits the neutron guides from 
the experiment into two arms, and each of them is dedicated 
to the analysis of one spin state. Two identical NANOSC detectors are installed at the end of these arms. Both detectors have a similar detection efficiency (at the percent 
level) as reported in Ref. \cite{Hel15}.

The data acquisition is carried out by the FASTER system (Section \ref{single_deteff}). The system time stamps all events and is able to measure the pulse charge 
with a maximum of 4 integration gates. The trigger is operating in a two 
dimensional space allowing the discrimination of pulses according to their amplitude and their duration (measured at the level of the amplitude threshold). Detailed investigations have shown 
that events with a pulse height larger than 3 mV (for an applied high voltage of about 800 V) and pulse width above 12 ns, at the 3 mV level, correspond to the best setting able to efficiently reject electronic noise as well as most of 
the \v{C}erenkov events (Section \ref{single_deteff}).

\subsubsection{Time distribution}
\label{Time_dist}

The nEDM measurement is performed from a large set of individual measurements called cycles. Each cycle is divided into 
three phases: the filling of the UCN storage chamber, the 
storage of UCN during which the neutron spin freely precesses and the emptying of the vessel. A switch located below the spectrometer 
is used to distribute the UCN flux from the source to the precession chamber (filling), then from the source to the detectors (monitoring) and finally from the 
precession chamber to the detectors (counting). The spin states of the UCN counted during the emptying sequence are analyzed and form the basis of 
the nEDM analysis.

\begin{figure}
\begin{center}
\resizebox{0.4\textwidth}{!}
{\includegraphics{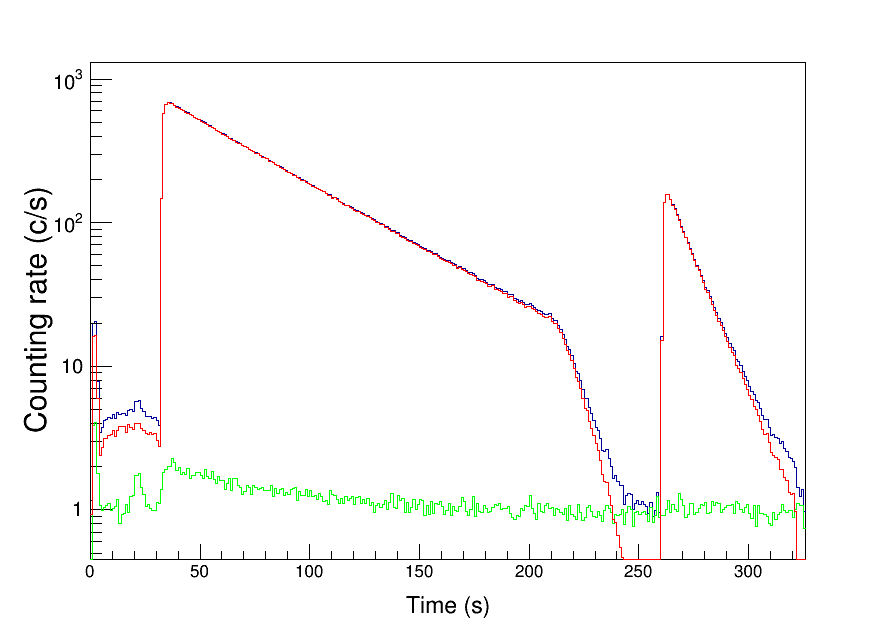}}
\captionsetup{width=0.8\textwidth}
\caption{\small Counting rate measured by the NANOSC detectors and plotted as a function of the time during a typical nEDM cycle (see text for details). 
The blue (dark grey) distribution corresponds 
to the total counting rate, the red one (intermediate grey), to the UCN rate and, the green one (light grey), to the background rate.}
\label{fig:time_dist}
\end{center}
\end{figure}

The time distributions of the background (gamma, electronic noise and \v{C}erenkov) and the UCN contributions are displayed in fig. \ref{fig:time_dist}. These distributions 
are normalized such that the vertical axis variable 
is a counting rate (counts/s). The first 34 s correspond to the filling of the precession chamber. During this phase, UCN leaking from the switch were counted. The background 
exhibited a small peak around 25 s. This was due to the activation of water present in the beamline from the magnets' cooling system circulating close to the experiment. 
During the storage phase (220 s), the UCN flux provided by the source was monitored. At large UCN rates, {\it i.e.} at the 
beginning of the monitoring sequence, the background was correlated to UCN detection (Section \ref{bckg_compo}). Finally, the storage chamber was 
emptied and the UCNs were detected for 60 s 
starting from 260 s. The average number of detected UCN per cycle was between 6000 and 7000 in 2014. No correlation between the background contribution 
and the UCN rate was observed during the counting sequence. 

\subsubsection{Charge distribution}
\label{Charge_dist}

\begin{figure}
\begin{center}
\resizebox{0.5\textwidth}{!}
{\includegraphics{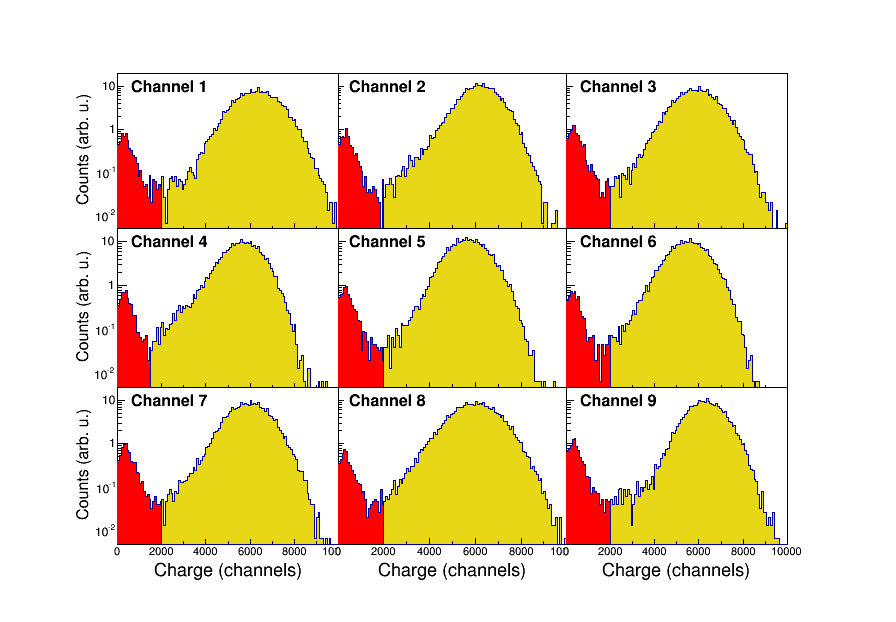}}
\captionsetup{width=0.8\textwidth}
\caption{\small Charge distributions measured by the nine channels of one NANOSC detector. The PMT high voltages were tuned in order to set all
UCN peaks at roughly the same location. A log scale is used for the vertical axis.}
\label{fig:NANOSC_charge_dist}
\end{center}
\end{figure}

The charge distributions of the nine channels of one NANOSC detector are shown in fig. \ref{fig:NANOSC_charge_dist}. The charge 
was collected from 9 ns 
up to 200 ns after the trigger which defines the starting time (t= 0 ns). Excluding the first 9 ns 
enabled better rejection of the 
the contribution of \v{C}erenkov events 
\cite{Ken11}. The full-energy peak associated with UCN detection is observed at large charges while the gamma contribution 
appear at low charges. The separation 
between both contributions 
is not as sharp as for the scintillator stack sample (fig. \ref{fig:molstick}). This is due to light-collection 
inefficiencies induced by the 80 mm long light guides used between 
the scintillators and the PMTs (no light guide was used with the former stack sample). The slight asymmetry 
of the UCN peak towards smaller charges is also in part explained by 
this light-collection deficiency.

\begin{table}[!htb]
\begin{center}
\captionsetup{width=0.8\textwidth}
\caption{\small Fraction of detected UCNs measured for each channel during the counting sequence of a 
standard nEDM run. The normalization is performed with respect to the total number of UCNs counted 
in each corresponding detector. The two numbers written below the channel label are the UCN counts 
fraction for each detector.} 
\vspace {0.3 cm}
\center
\begin{tabular}{|c|c|c|}
\hline
 \textbf{Channel 1} & \textbf{Channel 2} & \textbf{Channel 3} \\
 10.2$\%$ ; 10.1$\%$ & 11.7$\%$ ; 11.2$\%$  & 10.5$\%$ ; 10.8$\%$  \\
\hline
 \textbf{Channel 4} & \textbf{Channel 5} & \textbf{Channel 6} \\
 11.1$\%$ ; 11.3$\%$ & 12.7$\%$ ; 12.7$\%$ & 11.6$\%$ ; 11.9$\%$ \\
\hline
\textbf{Channel 7} & \textbf{Channel 8} & \textbf{Channel 9} \\
 10.3$\%$ ; 10.3$\%$ & 11.5$\%$ ; 11.0$\%$ & 10.3$\%$ ; 10.6$\%$ \\
\hline
\end{tabular}
\label{tab:rel_countrate_NA}
\end{center}
\end{table}

\subsubsection{UCN counting rate}
\label{UCN_rate}

\begin{sloppypar}
The number of UCNs was derived from the charge distributions: all events with a charge greater than a threshold ($T$) are counted as UCN. This threshold was set at the minimum 
of the counting rate between both contributions as illustrated in fig. \ref{fig:NANOSC_charge_dist} where the boundary between the gamma contribution in red 
(dark grey) and the UCN events 
in yellow (light grey) corresponds to the threshold location. The location of the minimum depends on the relative amount between the UCN contribution and the background 
contribution. It was defined during the emptying sequence {\it i.e.} for a moderate UCN counting rate. For high UCN counting rates, a lower charge threshold would be better suited. 
The increase of the background rate, observed at the beginning of the monitoring phase (fig. \ref{fig:time_dist}), is likely explained by this artefact. The influence of this threshold on 
the UCN counting rate is discussed in Section \ref{ng_discri}.  
\end{sloppypar}

For each detector, the distribution of the UCN flux over the nine channels was measured. It was nearly uniform over the 9 channels as shown in 
Table \ref{tab:rel_countrate_NA}. The surface of the scintillator array is slightly larger and not centered with respect to the exit Section of the 
simultaneous spin analyser UCN guides. These two features consistently explain the observed asymmetry between all channels.  

\begin{sloppypar}
In 2015, data were collected with a shorter duration for UCN monitoring (30 s). This allowed measuring the background contribution 
just before the emptying sequence. 
The total number of events measured during the emptying phase was then corrected for this background contribution in order to 
get another estimate of UCN rates. Both 
estimates are consistently below the 0.5 $\%$ level. 
\end{sloppypar}

\subsubsection{Low charge events composition}
\label{bckg_compo}

Extensive studies of the events located at small charges were performed. Two contributions were identified: gamma interactions and cross-talk between channels. 
The background rate amounted to 1--3 $\%$ of the total rate for the detectors. 

As discussed before, the gamma background contains two contributions: gamma interactions within the scintillators and the \v{C}erenkov light produced by gamma-induced 
electrons in the light guides. The use of very thin glasses reduces the amount of 
energy left by the gamma in the scintillator. As a result, these gamma events are mainly located in the small charge region \cite{Ban09}. The \v{C}erenkov 
rejection procedure benefits from the shortness of \v{C}erenkov pulses: the removal is performed by setting the pulse width threshold at 12 ns ({\it i.e.} only pulses with a duration 
larger than 12 ns at the 3 mV level are recorded) and by starting the charge integration 9 ns after the trigger (the short \v{C}erenkov pulses are therefore strongly truncated 
resulting in a low charge measurement).

The cross-talk between two neighbouring channels was reduced by wrapping the light guides and the scintillators edges with two PTFE layers (80 $\upmu$m thick). In addition, 
the amount of cross-talk events can be measured using the time stamping performed by the FASTER acquisition. Events detected in two neighbouring channels 
within a coincidence gate of 200 ns are considered as cross-talk events (except if their total charge corresponds to the charge of two simultaneously detected UCN). For each NANOSC 
detector, these cross-talk events correspond to 0.2--0.4 $\%$ of the total number of detected events. In addition, the amount of escaping light in the neighbouring channels is low. 
It lies in the low charge region and belongs to the background contribution. On the other hand, the amount of light in the channel of interest (i.e the channel from which the 
light is escaping) is slightly lower and enlarges the asymmetry of the UCN peak towards the low charge region. In any case, these events stay above the charge 
threshold and the number of UCN counts is not deteriorated by the cross-talk between channels.   

\subsubsection{Robustness of the neutron-gamma discrimination}
\label{ng_discri}

\begin{figure}
\begin{center}
\resizebox{0.45\textwidth}{!}
{\includegraphics{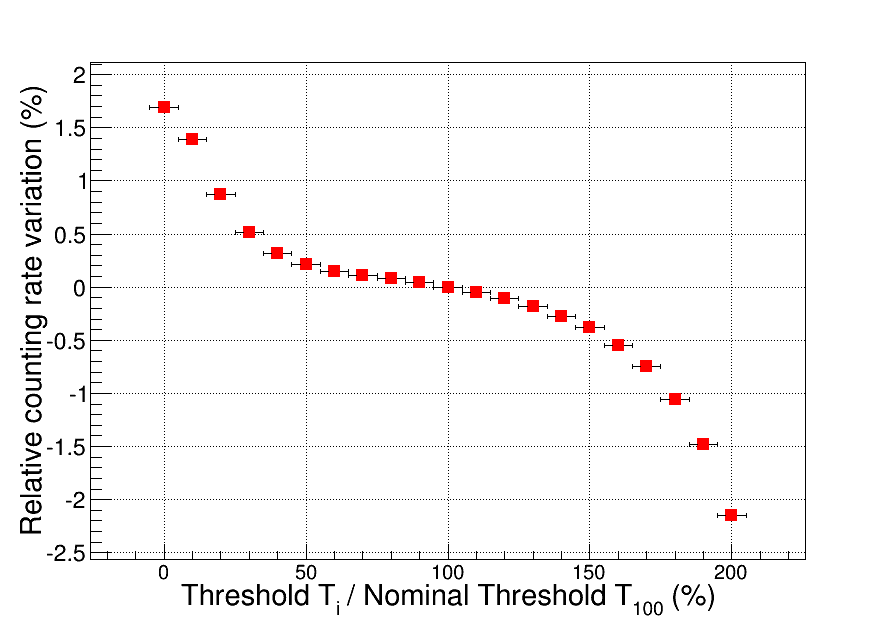}}
\captionsetup{width=0.8\textwidth}
\caption{\small Relative variation of the UCN counting rate displayed as a function of the charge threshold amplitude. On the horizontal axis, 
the threshold is relative to the nominal threshold $T_{100}$ 
and is expressed as a percentage. The vertical statistical error bars are smaller than the marker size.}
\label{fig:thresh_study}
\end{center}
\end{figure}

The influence of the charge threshold amplitude on the UCN counting rate is shown in fig. \ref{fig:thresh_study}. The study was performed during the 
counting time interval {\it i.e.} with the number of counted UCNs from which the nEDM analysis will be carried out. The number of UCNs, $N(T_{\rm{i}})$, was estimated 
for different charge thresholds $T_{\rm{i}}$ varying around the nominal threshold $T_{100}$ corresponding to $100$ on the horizontal axis of fig. \ref{fig:thresh_study}. 
This procedure was applied for each channel of the two detectors. Then, for each detector, the number of UCN was averaged over all channels and for all cycles 
of one run. The result is summarized in fig. \ref{fig:thresh_study} where the relative UCN counting rate variation, 

\begin{equation}
\Delta N_{\rm{i}}=\frac{<N(T_{\rm{i}})> - <N(T_{100})>}{<N(T_{100})>}
\label{ratio}
\end{equation}

\begin{sloppypar}
 is plotted as a function of relative charge threshold $(T_{\rm{i}} / T_{100})$. Around the nominal threshold, 
the counting rate variation was small and symmetric. 
Varying the threshold by 10$\%$ induced a UCN counting rate change of 0.15$\%$ ($\sim$ 5 counts over 3504) for the first detector and of 0.05$\%$ ($\sim$ 2 counts over 3531) 
for the second. The increase or decrease of the statistical error resulting from such a threshold change was around 0.07$\%$ and 0.02$\%$ for the two detectors. 
\end{sloppypar}

Any threshold variation can be interpreted as a variation of the charge measurement. For instance, the PMTs gain and the optical contact may evolve with time. This induces a variation of 
the charge measurement.
Such effects were monitored by measuring the UCN peak location as a function of time. For one month, variations up to 2$\%$ 
were observed among all channels. This corresponded to a counting rate variation of 0.01$\%$ (less than 1 count) and 0.025$\%$ ($\sim$ 1 count) for each detector. 
Any variation of the charge measurement at the few percent level therefore has a very small influence on the counting rate. The UCN counting procedure can thus be considered as robust 
and has been found not to influence the nEDM analysis.  

\section{Conclusions}
\label{conclu_gen}

We have presented a thorough study of UCN detection with several types of $^6$Li-doped glass scintillators. For all of them, the light 
readout is performed with PMTs. Single scintillators with a natural $^6$Li content (GS10) or with an enriched $^6$Li content (GS20) fully 
satisfied the UCN detection requirements. Thicknesses down to 100--200 $\upmu$m 
were manufactured in order to reduce both the gamma interaction probability and the amount of energy left by these gamma
interactions in the glasses. The GS10 scintillators are better suited than the GS20 scintillators for two main reasons: the Fermi potential (85 neV) as well as 
the edge events contribution are smaller. With a thickness of at least 150 $\upmu$m, GS10 scintillators present the same detection efficiency as 
a standard $^3$He gas detector. They are moreover able to handle counting rates up to 10$^6$ counts/s even if a specific treatment of pile-up 
events may be necessary.

A detection system based on stacks of two scintillators was investigated. The first layer is depleted in $^6$Li (GS30) while the 
second one is enriched in $^6$Li (GS20). The glass thickness was polished down to 60 $\upmu$m for the entrance glass (GS30) and to 120 $\upmu$m 
for the backing glass (GS20). The optical contact bonding was used between both scintillators to ensure 
a perfect optical contact. With this arrangement, most of the edge events can be suppressed resulting in an improved neutron-gamma 
discrimination.  

Based on these scintillator stacks, a detector was built for the nEDM experiment at PSI. It is made of nine 
channels and has been used for the nEDM experiment since 2010 without any major problem. The acquisition system (FASTER) is able to 
handle counting rates up to 10$^6$ counts/s. The system provides an event time stamp which allows for monitoring of the UCN detection 
over time during typical nEDM cycles. Depending on the UCN energy spectrum, the detector presents a detection efficiency similar or 
larger than that  
of a standard $^3$He gas detector. Finally, the UCN counting was performed via the measurement of the pulse charge. It was shown that the discrimination against any 
background contribution is robust and does not influence the nEDM data analysis.
Since 2014, two identical NANOSC detectors coupled to a new simultaneous spin analyzer have been operating in the nEDM experiment
\cite{Hel15}.

\section*{Acknowledgments}

\begin{sloppypar}
The NANOSC design couldn't be achieved without the mechanical team (D. Goupilli\`ere, H. de Pr\'eaumont, J. Lory, P. Desrue, C. Pain, B. Bougard 
and  Y. Merrer) and the technical staff (J. Bregeault, J.F. Cam, J. Harang, C. Vandamme) of LPC Caen who carefully manufactured the pieces. A special thanks 
to J.~Perronnel who have carefully handled the scintillators and to the FASTER team (D. Etasse, J. Hommet, B. Carniol, C. Fontbonne, T. Chaventr\'e) 
for their support. We would like to thank T. Brenner for his warm welcome 
and his support during the measurements at ILL and the PSI staff, in particular F. Burri and M. Meier, for their support at PSI. The LPC Caen and the LPSC acknowledge 
the support of the French Agence Nationale de la Recherche (ANR) under Reference No. ANR-09-BLAN-0046. This research was partly financed by the Fund for Scientific 
Research, Flanders; Grant No. GOA/2010/10 of KU Leuven; the 
Swiss National Science Foundation Projects No. 200020-144473 (PSI), No. 200021-126562 (PSI), No. 200020-149211 and 200020-162574 (ETH), and
Grants No. ST/K001329/1, No. ST/M003426/1, and No. ST/L006472/1 from the Science and Technology Facilities Council (STFC) of the United Kingdom. The 
original apparatus was funded by grants from the PPARC (now STFC) of the United Kingdom. Our Polish partners
wish to acknowledge support from the National Science Centre, Poland, under Grant No. UMO-2012/04/M/ ST2/00556.
\end{sloppypar}

%

\end{document}